\documentstyle[11pt,aaspp]{article}
\newcommand{\postscript}[2]
 {\setlength{\epsfxsize}{#2\hsize}
  \centerline{\epsfbox{#1}}}
\def\ref#1{\par\noindent \hangindent=0.4in \hangafter=1 #1 \par}
\def\eqalign#1{\null\,\vcenter{\openup\jot \m@th
  \ialign{\strut\hfill$\displaystyle{##}$&$
     \displaystyle{{}##}$\hfill \crcr#1\crcr}}\,}
\def\tempest%
{\begin{array}{ccc}
1 & 1 & 1 \\
1 & 1 & 1 \\
4 & 3 & 8
\end{array}}

\def\lsim{{{}_<\atop{}^{{}^\sim}}}
\def\cross{\rm cross}

\begin{document}

\title{Spectrophotometric Resolution of Stellar Atmospheres with 
Microlensing}

\author{B. Scott Gaudi, Andrew Gould\footnote[1]{Alfred P.\ Sloan Foundation Fellow}}

\affil{Ohio State University, Department of Astronomy, Columbus, OH 43210 \\
gaudi,gould@astronomy.ohio-state.edu}

\begin{abstract}

Microlensing is  a powerful tool for studying stellar atmospheres
because as the source crosses regions of formally infinite
magnification (caustics) the surface of the star is resolved, thereby
allowing one to measure the radial intensity profile, both
photometrically and spectroscopically. However, caustic crossing
events are relatively rare, and monitoring them requires intensive 
application of telescope resources.  It is therefore essential that the
observational parameters needed to accurately measure 
the intensity profile are quantified.  We calculate the expected 
errors in the recovered radial intensity profile as a function of the 
unlensed flux, source radius,
spatial resolution the recovered intensity
profile, and caustic crossing time for the two principle
types of caustics: point-mass and binary lenses.  
We demonstrate that for both cases
there exist simple scaling relations between these 
parameters and the resultant errors.  We find that the error as a
function of the spatial resolution of the recovered
profile, parameterized by the number of radial bins, increases as $N_R^{3/2}$, 
considerably faster than the naive $N_R^{1/2}$ expectation.  
Finally, we discuss the relative advantages of binary caustic-crossing 
events and point-lens events.  Binary events are more common, easier to
plan for, and provide more homogeneous information about the stellar
atmosphere.  However, a sub-class of point-mass events with low impact
parameters can provide dramatically more information provided that they can
be recognized in time to initiate observations.

\end{abstract}

\centerline{submitted to {\it The Astrophysical Journal}: Feb 16, 1998}
\centerline{Preprint: OSU-TA-2/98}

\newpage
\section{Introduction}

Originally proposed by Paczy\'nski (1986) as a method to detect the
presence of massive compact objects in the halo of our Galaxy,
microlensing has increasingly been recognized as a 
tool for studying a broad range of astrophysical phenomena.  Various 
applications include detection and characterization of binary and
planetary systems (Mao \& Paczy\'nski 1991; Gould \& Loeb 1992), 
reconstruction of the stellar mass function down to masses below the 
hydrogen burning limit (Paczy\'nski 1991; Griest et al.\ 1991; Gould
1996), measurement of the rotation speed of giants (Gould 1997), 
measuring the transverse velocity of galaxies (Gould 1995a), and
probing the central engines of quasars (Gould \& Gaudi 1997).  
Recently, Valls-Gabaud (1994, 1997) and Sasselov (1996) have proposed 
using microlensing to study stellar atmospheres.  
Here we analyze this application in detail.

Currently, three collaborations (MACHO, Alcock et al.\ 1997a; OGLE, 
Udalski et al.\ 1997; EROS, Ansari et al.\ 1996) have ongoing projects
that survey the Galactic bulge with roughly nightly sampling in order
to detect microlensing events.  Over 60 events per year are being
detected.  These data are being analyzed real time which
has allowed
MACHO to issue ``alerts'', notifications of ongoing events detected
before the peak.  Two follow-up collaborations (PLANET, Albrow et al.\
1997, 1998a; GMAN, Alcock et al.\ 1997b) have formed
in order to
monitor these alerts around the clock with high precision and high
temporal sampling with the aim of detecting (and further alerting on)
deviations from the standard microlensing light curve, 
such as would be expected from binary lenses, binary sources, finite
sources, and parallax.  These deviations are useful in that they can
provide additional information about the lens and/or
source.  In addition, a collaboration has formed to conduct
{\it spectroscopic} monitoring of alerted events in order to study
the source stars in detail and has observed several events to date
(Lennon et al.\ 1997).  Thus, the prospect for
the real-time detection and monitoring of light curve anomalies both
photometrically and spectroscopically is promising.

All gravitational microlenses have caustics, defined as the set of
points in the source plane where the magnification of a point source is formally
infinite.  When a finite source crosses a caustic, the gradient
of the magnification over the source is large, and therefore different parts of the source are 
magnified by different amounts.  Hence, the source is partially resolved.  Different parts of 
the source are resolved at different times during the caustic crossing, and thus, by taking a 
series of measurements during the course of the crossing, one can recover the intensity profile
of the source.  Several workers realized that finite source effects could be useful for
breaking or partially breaking the degeneracy among microlensing event parameters 
(Gould 1994; Nemiroff \& Wickramasinghe 1994; Witt \& Mao 1994; Maoz \& Gould 1994) 
and that variations  in the surface profile could be exploited to this
end (Witt 1995; Loeb \& Sasselov 1995; 
Gould \& Welch 1996).  However, Valls-Gabaud (1994, 1997), Sasselov (1996), and Heyrovsky, 
Sasselov, \& Loeb (1998) proposed to exploit the same effects for radically different aims.
Instead of using finite-source effects to learn more about the lens, they
sought to learn more about the source.  The basic idea is as follows:
Imagine that one could image separately different annular rings on the
surface of a star.  In effect one would be sampling different depths
of the photosphere.  Since the temperature varies as a function of
depth, the broad spectral energy distribution would change with
annulus, with more blue light near the center (greater depths) and
more red light near the outer limb (lesser depths).  That is, the star
would be limb-darkened, and more in the blue than the red.  Since
different spectral lines form at different depths, one would expect
that the detailed spectral profile would vary as a function of annular
radius.  Hence, the entire atmosphere could be studied as a function
of depth by resolving the two-dimensional (radius and wavelength)
spectral profile of the star.  Currently, it is only possible 
to study stellar atmospheres in this way for the Sun and eclipsing
binaries. However, since the surface of the source star is partially resolved
during a microlensing caustic crossing, one can 
also probe the atmospheres of 
the source stars for these types of events.

Both Valls-Gabaud (1994, 1997) and Sasselov (1996) used specific
stellar atmosphere models to construct broad-band and spectral
line brightness profiles, and then used these profiles to predict in
detail the variations in the broad-band color or equivalent width of specific
lines that one would expect during the course of a point-mass microlensing
event.  Sasselov (1996), and Heyrovsky et al.\ (1997) also
consider the effects of star spots 
on the microlensing light curve. In
addition, they compared
their predictions to MACHO Alert event 95-30, a point-mass
caustic crossing event for which spectra were taken during the course
of the crossing, and for which variations in the optical TiO bands were detected
(Alcock et al.\ 1997b).  All of these authors predict that the
color and spectral-line variations during the caustic crossing should
be significant and note that this provides an entirely new method of
studying stellar atmospheres.

Although caustic crossing events are in principle useful for studying the
atmospheres of stars, these events are rare, and they typically last for only about 
7 hours (for a giant source).  For
this method to be successful, it 
is essential that observers have a clear sense of what can be
accomplished with these events, since substantial telescope resources
are likely to be expended.
To this end, we approach this topic from
another perspective.  We quantify the intrinsic ability of
both point-mass and binary lens microlensing caustic crossings to 
recover the radial variation of the intensity of the star for any
arbitrary wavelength, and hence, for any
spectral line.  Specifically, we calculate the fractional error in the
recovered intensity profile as a function of the unlensed flux of the
source, the duration of the measurements, the size of the telescope, 
the magnification of the event, and the 
spatial resolution of the
recovered intensity profile. This information will be useful to 
observers in making rational decisions about which events to follow,
and what resources are required to address specific questions.

\section{Formalism}

\subsection{Lens Geometry, Magnification, and Caustics}

Microlenses come mostly in two flavors: point lenses and binary lenses.  The magnification 
structure of these two types of lenses is essentially different, so we consider them separately.

The flux of a point source being microlensed by a point lens, $F$, 
can be expressed in terms of
five parameters, $F_0$, the unlensed flux of the source, $B$,
the flux of any unresolved background light, $t_0$, the time of maximum
magnification, $\beta$, the impact parameter of the event, and $t_e$,
the time scale 
of the event.  These are related by $F=F_0A(t)+B$,
where $A(t)$ is the magnification, which is itself a function of
the parameters $t_0$, $t_e$, and $\beta$, and is given by, 
$$
A[x(t)]= {x^2+2 \over {x(x^2+4)^{1/2}}}, \,\,\,\,\, 
x(t)=\biggl[\beta^2 +{(t-t_0)^2\over t_e^2}\biggr]^{1/2}.
\eqno(2.1)
$$
Here $x$ and $\beta$ are in units of the Einstein ring radius, $R_e$, given by,
$$
R_e= {\left[ 4GM D_{OL}(1-D_{OL}/D_{OS})\right]^{1/2} \over c^2},
\eqno(2.2)
$$
where $M$ is the mass of the lens, and $D_{OL}$ and $D_{OS}$ are the distances to the lens and 
source respectively.  The time scale 
is related to the Einstein ring radius by $t_e=R_e/v$,
where $v$ is the transverse speed of the lens relative to the observer-source 
line of sight.  Note that for $x \ll 1$, $A(x)\simeq x^{-1}$, and 
that the magnification therefore diverges for a point
source.  Thus the point $x=0$ corresponds to the caustic for 
the point-lens case.  
Consider a source of radius $R_*$.  The radius of this star, projected onto the lens
plane, in units of $R_e$ is given by,

$$
\rho = {R_* \over R_e} {D_{OL}\over D_{OS}}.
\eqno(2.3)
$$
For
$M=0.3\,M_\odot$, $D_{OL}/D_{OS}=0.75$, $D_{OS}=8\,\rm kpc$, and $R_*=10\,
R_\odot$, $\rho \simeq 0.02$.  For a source of this size, equation (2.1)
remains valid for the majority
of the event.  However, when the lens comes within $\sim 2\rho$ of the source, i.e., for
events with $\beta \lsim 2\rho$, the magnification deviates from
equation (2.1), and the finite size of the source must be considered.  The 
time scale for this deviation is roughly the
crossing time for the source, $t_{\cross}=\rho t_e \simeq 7(R_*/10
R_\odot)\,{\rm hours}$ for typical bulge events. For these
events, the source parameter $\rho$ enters into the calculation of the 
magnification.  The probability of such an event is $\sim \langle \rho\rangle$, where
$\langle \rho\rangle$ is the average radius of the source stars being monitored.
If $100$ giant-source events were detected during a year, 
then the lens would transit the source for about 5 of them and these would
exhibit finite-source deviations (Gould 1995b). 

The magnification of a point source being lensed by a binary depends on
the same parameters as the point lens, $t_0$, $t_e$, and $\beta$,
along with the additional parameters $b$, the separation
of the lenses in units of $R_e$, $q$, the mass ratio of the lenses,
and $\theta$, the angle between the axis of the binary and the
trajectory.  Unfortunately, the dependence of the magnification on these
parameters in the general case has no analytic form.
Since the method of determining the magnification in the general case
has been described elsewhere (see, e.g.\ Witt \& Mao 1995), and since we are only interested in
caustic crossings, we will 
employ a simplified formalism for these crossings.  For nearly
equal mass binary lenses, the caustics are composed of curved lines (called folds) 
whose extent is of order the Einstein ring radius.  
Thus the probability of encountering a fold caustic during a binary-lens
event is almost unity.  For a point source, the excess magnification near the
fold caustic is approximately $A\propto x^{-1/2}$ (Schneider, Ehlers,
\& Falco 1992), where $x$ is the distance from the caustic, and the
magnification diverges at $x=0$ (at
the caustic).  For a finite source of radius $\rho$, the magnification deviates from
this form for distances $x\lsim 2\rho$, and the finite size of the
source must be considered.  Binary lenses also have cusps, points
where two fold caustics merge, and the magnification structure near a
fold is quite different from that near a cusp. In particular, it
cannot be described by the same equations.  While
the probability of encountering a fold caustic is nearly unity, the probability of
encountering a cusp is $\sim N\rho$, where $N$ is the number of cusps.  For typical binary
lenses, $N=6$, and thus the probability is $\sim 20\%$.  If $100$
events were discovered towards the bulge per year, and 
$5\%$ of these were binaries, we would expect only
one cusp crossing
per year, while we would expect $5$ events with fold crossings.  Hence,
we will consider only fold crossings here.

Before continuing to the next section, where we consider the
magnification one expects when a finite source crosses
the two types of caustics considered above, we include a brief
discussion concerning notation. Note that there are two characteristic scales in
this problem, $R_e$ and $\rho$, which are related by equation (2.3).  
Either of these could be used as our fiducial scale.  For
definiteness, we choose to scale all quantities by $R_e$.  For the remainder of
the discussion,  we will also assume that $t=0$ is the time when the
center of the source crosses (or comes closest to) the caustic. For
the point-mass case, this simply means setting $t_0=0$.   Finally, for
the binary lens, the caustic crossings come in pairs.  Throughout we
will be referring to the second caustic crossing (since it is this
crossing that can be predicted, see \S\ 5), when the source is
moving from the inside of the caustic structure to the outside.

\subsection{Extended Sources}

The magnification of an extended source is given by:
$$
A_{es}(t)= {{\int d^2 r\, A(t;{\bf{r}})I({\bf{r}})}\over
{\int d^2 r\,I({\bf{r}})}},
\eqno(2.4)
$$
where $I({\bf{r}})$ is the intensity profile of the source,
$A$ is the magnification of a point source at $\bf r$, and the integral is over 
the area of the source.  The numerator and denominator are respectively the lensed and 
unlensed flux of the source.  Assuming that the intensity profile has azimuthal symmetry,
and using polar coordinates, this becomes,
$$
A_{es}(t)={{\int_0^{\rho} d r\,r I(r) {\cal{A}}(t;r)} \over {\int_0^{\rho} d r\,
r I(r)}},
\eqno(2.5)
$$
where ${\cal{A}}(r)$ is the angle averaged magnification function,
$$
{\cal A}(r)\equiv {1\over 2\pi}\int_0^{2\pi} A(r,\theta)d\theta.
\eqno(2.6)
$$ 
This definition is useful because the magnification geometry of the 
lens is entirely isolated in this function.  Once 
the function has been calculated,
it can be convolved with any source intensity profile to give the total magnification
$A_{es}$.  Furthermore, it is the shape of this function that determines how
well one can resolve the source, i.e. if ${\cal A}$ is highly peaked at a particular
value of $r$, then the majority of the lensed flux will be coming from a small
range of radii near
$r$, and therefore the lens is resolving the source.

\subsection{Angle Averaged Magnification Functions}

For both fold and point caustic crossings, the approximate form of the
function ${\cal A}$ can be calculated analytically. 
When a source crosses a fold caustic, two images appear or disappear.  For typical
total binary-lens mass, the 
size of the source, $\rho$, is considerably smaller
than the Einstein ring radius of the lens, $\rho \ll 1$, and thus 
the magnification of the other three images changes very little as the source crosses
the caustic.  Also, the curvature of the caustic is typically very small across
the source.  In this regime, the magnification in the vicinity of the fold caustic
can be approximated by (Schneider et al.\ 1992),
$$
A(x)=A_0 + \left(b_0 \over R_e\right)^{1/2} x^{-1/2},\,\,\,\,\, x>0\,\,\,\,\, {\rm (inside\,\, the\,\, caustic)}
\eqno(2.7)
$$

where $A_0$ is the total magnification of the three unrelated images, $x$ is
the distance to the caustic, and $b_0$ describes the scale of the caustic.  For
approximately equal mass binaries, $b_0 \sim R_e$.  For $x<0$ (outside
the caustic), $A(x)=A_0$.  Defining
 $z=x/r$, where $x$ is now the distance from the center of the source to the caustic,
and setting $b_0=R_e$, the angle-averaged magnification function is,
$$
{\cal A}(r)=A_0 + r^{-1/2}{\it{j}}(z),
\eqno(2.8)
$$
where,
$$
{\it{j}}(z)= \cases{ 
0   & $z<-1$   \cr
{2^{1/2}\over \pi}K\left[ \left({1+z\over 2}\right)^{1/2}\right]&$|z|<1$ \cr
{2\over \pi}(1+z)^{-1/2} K\left[ \left({2\over 1+z}\right)^{1/2}\right]& $z>1$ \cr
}.
\eqno(2.9)
$$
Here $K$ is the complete elliptic integral of the first kind.  
Figure 1 shows ${\cal A}(r/\rho)$ versus $r/\rho$ for $A_0=0$, 
 $\rho=1$, and $x=-t/t_e=z\rho=5/3\rho$, 
$\rho$, $2/3\rho$, $1/3\rho$,
 $0$, $-1/3\rho$, and $-2/3\rho$.  For
 $x=-t/t_e > \rho$ 
(when the source is entirely contained within the caustic), 
the gradient of the magnification
across the face of the source is small, and furthermore 
${\cal A}(r/\rho)$ has no maximum.  This
implies that these times are not useful 
for resolving the source.  For $0<x=-t/t_e<\rho$, 
 ${\cal A}(r/\rho)$ shows a sharp peak at  $r=x$, 
and thus the source is being resolved.  However, other annuli are
being significantly 
magnified and are thus contributing significantly to
the total light, and therefore the resolution will be degraded
somewhat. By contrast, for $-\rho <x=-t/t_e<0$, 
and in the limit 
  $A_0\rightarrow 0$, only those annuli that have just crossed the caustic
contribute to the total light.  It is therefore these times that are
most useful for resolving the source.

For a source of uniform brightness, the total magnification is
(Schneider \& Weiss 1987),
$$
A_{es}(r)=A_0+ r^{-1/2}{\cal{J}}(z),
\eqno(2.10)
$$
where
$$
{\cal{J}}(z)=\cases{
0  & $z <-1$ \cr
{2^{5/2}\over3\pi} \left( (1-z) K\left[ \left({1+z \over 2}\right)^{1/2}\right] +2z E\left[\left({1+z}\over 2\right)^{1/2}\right] \right)& $|z|<1$\cr
{8 \over 3\pi}(1+z)^{1/2} \left(z E\left[\left(2 \over 1+z \right)^{1/2}\right] -(z-1)K\left[\left(2\over {1+z}\right)^{1/2}\right]\right)& $ z> 1$\cr
},
\eqno(2.11)
$$
and $E$ is the complete elliptic integral of the second kind.
Note that in equations (2.8) and (2.10), the dependences on $z$ and $r$
are separable for $A_0=0$.  This implies that, 
for a fixed value of
 $z$, both ${\cal A}$ and $A_{es}$ scale simply as $\rho^{-1/2}$. 

We now turn to the point lens case.  When the separation between the lens and the
source is much smaller than the Einstein ring radius, $x \ll R_e$, the magnification
can be approximated by,
$$
A(x)= x^{-1}.
\eqno(2.12)
$$
The angle-averaged magnification function is then,
$$
{\cal A}(r)=r^{-1}{\it{b}}(z),
\eqno(2.13)
$$
where,
$$
{\it{b}}(z)= {2\over\pi}(1+z)^{-1}K\left[ {4z\over (1+z)^2}\right],
\eqno(2.14)
$$
and $K$ is the complete elliptic integral of the first kind.  Figure 2 shows
 ${\cal A}(r/\rho)$ versus $r/\rho$ for $\rho=1$, and trajectories with 
 $x=(t^2/t_e^2+\beta^2)^{1/2}$, for 
impact parameters $\beta=0$ (solid curves) and 
 $\beta=0.5\rho$ (dashed curves), and $t/t_e=-\rho$, $\,-2/3\rho$,
 $\,-1/3\rho$, $\,0$, $\,1/3\rho$, $\,2/3\rho$, and $\,\rho$. 
Here ${\cal{A}}$ achieves a local maximum whenever $r = x$, implying that
these radii are being partially resolved at these times.  Unfortunately,
for any impact parameter $\beta >0$, there is a range of source radii
at which ${\cal{A}}$ is never at a maximum, those for which
 $r<\beta$.  These radii are not resolved during the caustic crossing,
and thus very little information about the radial intensity profile
will be gained for this range of source radii.

The total magnification for a source of uniform brightness is
(Schneider et al.\ 1992),
$$
A_{es}(r)=r^{-1}{\cal{B}}(z),
\eqno(2.15)
$$
where
$$
{\cal{B}}(z)=\cases{
{4\over \pi}E(z) & $z \le 1$\cr
{4\over \pi}z  \left[ E(1/z)- (1- z^{-2}) K(1/z) \right] & $ z\ge 1$\cr
}.
\eqno(2.16)
$$
As in the fold case, the dependences on $z$ and $r$ in equations
(2.13) and (2.15) are separable.  Thus, for a fixed value of $z$, 
 ${\cal{A}}$ and $A_{es}$ scale as $\rho^{-1}$.

\section{Error Analysis}
The flux measured from a star of finite size being lensed is 
$F(r)=A_{es}F_0+B$, where the
general form of $A_{es}$ is given in equation (2.4).  For most caustic crossing events,
the magnification is high, and $A_{es}F_0 >> B$, thus we will henceforth use the approximation
that $B=0$.  Consider a star divided up into $N_r$ radial bins, each located at $r_i$. 
Defining $\Delta r_i$ as the width of this bin, 
the flux of the star is given by,
$$
F(t)=\sum_i^{N_r} 2 {\cal A}(r_i) I(r_i) r_i \Delta r_i,
\eqno(3.1)
$$
where $\cal A$ is the angle averaged magnification in bin $i$.  For
the case of a large number of bins, or $\Delta r_i \ll \rho$, the
magnification in each bin is well approximated by equations (2.8) and (2.13).  For
a small number of bins, however, these forms are not good
approximations, and one should use the total magnification in each
bin, which can be obtained by either integrating equations (2.8) and
(2.13) over
the radii of the bin, or, equivalently, by using equations (2.10) and (2.15), and
subtracting the flux within the 
inner  
radius of the bin from the flux within the
outer
 radius of the bin, and dividing by the area of the annulus.  
In practice, the latter is simpler.  The parameters one would
like to recover are $I(r_i)$, the 
mean intensity in bin $i$ for a
certain wavelength.  In order to do this, however, one must first know
 ${\cal{A}}(r_i)$, which implies knowing the parameters $\rho$,
 $\beta$, $t_e$, and $t_0$ for the 
point-lens case, and the additional
parameters $\theta$, $b$, and $q$ for the binary case.  In principle,
one could determine these parameters entirely within the context of
the spectrophotometric measurements. However, variations in the 
seeing and extinction make this difficult.  Fortunately, this is not a significant
hindrance because independent broad-band measurements can be used to
constrain these parameters.  Since the follow-up
collaborations already monitor these events, and, in one case, have
already used broad-band measurements to constrain these parameters
{\it and} measure the limb-darkening of the source star (Albrow et
al.\ 1998b), this should not be a problem in practice, although
it does imply that the spectroscopic and photometric follow-up
collaborations should closely coordinate their efforts.
Now suppose that a series of spectrophotometric measurements
 $F(t_k)$ are made at times $t_k$, with uncertainties $\sigma_k$, and
these  measurements are 
fitted to equation (3.1).  The parameters of this
fit are $I_i$, and the covariance
matrix of the errors in these parameters is given by $c_{ij}$, where
$$
c=b^{-1},\,\,\,\,\,\, b_{ij}=\sum_k \sigma_k^{-2} {\partial F(t_k)\over \partial I_i}
{\partial F(t_k) \over \partial I_j},
\eqno(3.2)
$$
and $\partial F(t_k)/ \partial I_i = 2 {\cal A}(r_i) r_i \Delta r_i$.
The variances in the parameters are then the
diagonal elements of $c_{ij}$, and thus the
fractional error in each parameter is 
$\delta I_i/I_i=(c_{ii})^{1/2}/I_i$.  Assuming photon-limited precision,
the fractional errors scale as the 
square root of the total number of photons received, which in turn scales
as $(A_{es} F_0)^{1/2}$.  From equation (2.10), $A_{es}\propto \rho^{-1}$, 
for the point caustic, and from equation (2.15),
$A_{es} \propto \rho^{-1/2}$ for the fold caustic.
Let $\gamma$ be the rate at which a telescope
receives photons from the unmagnified star
in a certain spectral range, and assume that measurements are made
continuously during a source-radius crossing
time $t_{\cross}$.  
Then the square root of the total number of photons scales
as $\rho^{-\nu}(\gamma t_{\cross})^{1/2}$, where $\nu=1/2$ for the point caustic and 
 $\nu=1/4$ for the fold caustic.  The fractional errors will also
depend on the spatial resolution of the recovered intensity profile,
 $N_r$. Assuming Poisson statistics and no correlations between
the parameters $I_i$,  one would expect the fractional errors in $I_i$
to scale simply as $N_r^{1/2}$.   However, we find that the
parameters $I_i$  are correlated, and the errors in fact scale approximately 
as $N_r^{3/2}$.
We justify this assertion in the next section.
Taking these scalings out of $c_{ii}$, we have
$$
{\delta I_i \over I_i}= N_r^{3/2} (\gamma t_{\cross})^{-1/2} \rho^{\nu} ({\tilde c}_{i,i})^{1/2},
\eqno(3.3)
$$
where $({\tilde c}_{ii})^{1/2}$ is the normalized fractional error for
 $N_r=1$, $\rho=1$, $\gamma t_{\cross} =1$. 
The specific form of $({\tilde c}_{i,i})^{1/2}$ will depend on the details of
the caustic encounter
and will therefore vary on an event-by-event
basis.  In the next section, we consider specific cases, and evaluate
 $({\tilde c}_{ii})^{1/2}$ directly.

\section{Results}

\subsection{Scaling with Number of Radial Bins}

In order to justify the assertion that the errors scale as $N_r^{3/2}$, we consider a specific
example for each caustic case.  For the point case, we consider an event with impact parameter
 $\beta=0$, and we assume that this event is observed from $t/t_e=0$ 
to $t/t_e=\rho$, i.e. for one source crossing time.  We then evaluate the normalized error,
 $({\tilde c}_{i,i})^{1/2}$ for this event as a function $r_i$, the radius of the bin, 
for different values of $N_r$.  This is shown in Figure 3a, for
 $N_r=10,20,40$.  If the errors scaled exactly as $N_r^{3/2}$, these curves
would be indistinguishable. In fact, the normalized errors decrease
slightly for larger $N_r$, indicating that the errors do not increase
quite as fast as $N_r^{3/2}$, although this depends
on $r_i$. For the fold case, we assume that the event is observed from
$x= -t/t_e=0$ when the center crosses the caustic until
 $x=-t/t_e=-\rho$ when the source moves completely outside the caustic.
In Figure 3b, we show $({\tilde c}_{ii})^{1/2}$
for $N_r=10,20,40$.  Here the scaling is nearly perfect.  We will henceforth assume that the
scaling is perfect for both cases, and therefore that
${\tilde c}_{ii}$ 
has no dependence on $N_r$.

\subsection{Fractional Errors}

In order to
evaluate the fractional errors explicitly, we adopt
some fiducial parameters.  For a
giant in the bulge, and a typical microlensing event, $\rho\sim 0.02$,
and we assume that observations are taken continuously for a
time, $t_{\cross}\sim 7\, {\rm hours}$.  We 
use $N_r=10$ for our fiducial number of radial bins, and, for the
moment, we assume a uniformly bright source, as this is the most
general way to quantify the error in a model-independent way.  In the
next section, we consider a source with limb darkening.  
The fractional 
error in the intensity profile for these parameters is,
$$
{\delta I_i \over I_i}= \left({\delta I_i \over I_i}\right)_0 \left({\rho\over0.02}\right)^{\nu}
\left({t_{\cross}\over 7\, {\rm hr}}\right)^{-1/2}
 \left({\gamma\over0.4\, {\rm{s^{-1}}}}\right)^{-1/2} \left(
{N_r\over 10}\right)^{3/2}.
\eqno(4.1)
$$
where $\nu=1/2$ for the point caustic and $\nu=1/4$ for the fold
caustic, and $\gamma=0.4\,{\rm{s^{-1}}}$ 
is approximately the flux of photons per spectral resolution element
from a star with $V=17$ that can be acquired with a 2m telescope
and a spectrograph with $1\,{\rm \AA}$ resolution.   In Figure 4a 
we show $(\delta I/I)_0$ as a function of $r_i$ for the point-lens
case, for $t/t_e=0$ to $\rho$, and  for four different impact
parameters, $\beta/\rho=0, 0.25, 0.50$, and $0.75$
(solid, dotted, short dashed and long dashed lines).
It is clear
from this diagram, that for $r > \beta$, the fractional
error is small, $\lsim 10\%$.  One could, in principle, improve the
fractional error by a factor $2^{1/2}$ by observing from 
 $-\rho <t/t_e < \rho$, i.e. for two source crossings.  However, this
will prove difficult in practice (see \S\ 5).  While the errors are
small for  $r > \beta$, there is almost no information
for radii $r < \beta$, as expected, since the lens is not passing
over this region of the star. In Figure 4b we
show $(\delta I/I)_0$ as a function of $r_i$ for the fold caustic
with $A_0=0$, and four different cases.  The short dashed line corresponds to
measurements taken from $x=-t/t_e=\rho$ to $x=0$, 
one crossing time
from when the edge of the source 
first crosses the caustic to when the source is exactly centered on the
caustic, i.e., the first half of the caustic crossing.  The 
solid line is for $x=0$ to $x=-t/t_e=-\rho$ 
(the second half of the caustic crossing); the 
dotted line is for $x=-t/t_e=\rho$ to $x=-t/t_e=-\rho$ 
(the entire caustic crossing), and
the long-dashed line is for $x=-t/t_e=2\rho$ to $x=-t/t_e=-\rho$.  
 From this figure one learns three 
things.  First, the errors are reasonable, $\delta I /I \lsim 20\%$ for typical parameters.  
Second, times immediately after the center of the source crosses
the caustic are the most crucial for recovering the
intensity profile accurately. Finally, it is only the times just before until just after the
caustic crossing that are useful for resolving the source.  We also
show the errors for the case that $A_0=1$ (dashed-dot line).
Clearly, the magnification of the images not associated with the caustic 
does not greatly affect the resultant errors.

\subsection{Effects of Limb Darkening}

In this section we examine the effect limb darkening has on the
fractional error in the recovered intensity profile.  We can
anticipate that limb darkening will serve to increase the errors in the outer annuli
relative the constant surface-brightness 
case because there will be fewer
photons coming from outer annuli, and thus the errors will be larger.
For the inner annuli, the fractional errors will be smaller compared
to the constant surface-brightness case, because
the source is, in essence, more compact, and thus the net
magnification will be larger, and therefore the errors smaller.  

To quantify this effect, we apply the same formalism as in \S\ 4.3,
except that
we adopt the following parameterization of the surface
brightness as a function of radius,
$$
{I(r)\over I(0)}=1 - \kappa_1 Y - \kappa_2 Y^2, \,\,\,\,\,
Y=1-\left(1-{r^2\over\rho^2}\right)^{1/2},
\eqno(4.2)
$$
where $\kappa_1$ and $\kappa_2$ are the limb-darkening coefficients.
We adopt the coefficients for a cool (4500 K) giant ($\log{g}=1.5$)
of solar metallicity from Manduca, Bell, \& Gustafsson (1977) and
Manduca (1979).  These are $\kappa_1=0.798,0.567,0.139$ and 
 $\kappa_2=-0.007,0.114,0.259$ for the V, I and K bands, respectively.
Since stars are less limb darkened in the infrared, 
the results for the K band will be the most similar to the uniform
surface brightness case, 
while those for V will be least similar.
The results for the same fiducial parameters used
in \S\ 4.2
($\rho=0.02$, $t_{\cross}=7\,{\rm hours}$, 
 $\gamma=0.4\, {\rm s^{-1}}$, $N_r=10$) are shown in Figure 5
along with the uniform surface-brightness 
case.  For the point-caustic
case [panel (a)], we have used a trajectory with $\beta=0$ and 
assumed that observations are taken for 
$0 < t/t_e <\rho$. 
For larger values of $\beta$, the fractional 
errors will deviate more dramatically from the uniform surface
brightness case, since for large values of $\beta$, only the outer
annuli are being effectively resolved, and it is these annuli that are
affected most by limb darkening.  For the fold caustic case [panel (b)], 
we have assumed  
$-\rho < x=-t/t_e < \rho$.  
As expected, for both the point-caustic and fold-caustic 
cases, the fractional errors are somewhat larger than 
for the uniform source at larger radii, and somewhat smaller at smaller
radii.  Also, the differences between the uniform source and 
limb-darkened source
decrease for longer wavelengths.  In general, however, for
both the point and fold caustics, the differences between uniform
and limb-darkened sources is modest.  For fold caustics the effect 
is $\lsim 25\%$ over the entire
range of radii.  For point caustics the effect is small over most of the
star but the error can more than double at the very limb of the star when
observed in $V$ band, the bluest color considered here.

\section{Discussion}

        While both point-mass lenses and binary lenses
can in principle be
used to resolve the two-dimensional (radial + spectral) profile of a star,
binary lenses are substantially easier to use.
First, for a binary one
always has warning of the second caustic crossing.  When the source crosses
the caustic the first time it is suddenly magnified by a factor
$\rho^{-1/2}\sim 7$ and hence is easily recognized.  The second crossing
can then be expected in several days to several weeks.  Intensive photometric
monitoring (now routinely undertaken by PLANET and GMAN) can then be analyzed
to make a more precise prediction.  From Figure 4b it is clear that the
most useful portion of the second caustic crossing is the final $\sim 70\%$
of the time that the source actually straddles the caustic.  The onset
of this optimal period can be judged extremely accurately if photometric
monitoring is proceeding simultaneously, and reasonably accurately even
one day in advance.  By contrast, there is no way to guarantee a priori
that a point-mass caustic crossing will occur because one does not know
the size of the Einstein ring projected onto the source plane beforehand, and
hence one does not know $\rho$.  Using optical photometry alone, one can
``predict'' a source crossing only at about the time it begins.  Using
optical/infrared photometry, it could be predicted at $r\sim 1.5\,\rho$
(Gould \& Welch 1995), but this would leave only a few hours' warning.

Second, fold caustics generically provide information about the
entire radial profile of the star while point caustics provide information
only for annuli of the star that are greater than the impact parameter, 
$\beta$.  See Figure 4.  Again, it is virtually impossible to predict in 
advance for which events $\beta\ll \rho$ (and hence for which 
events one can resolve
essentially the whole star), although once the peak occurred, 
these events could
be recognized provided that the star 
was being monitored photometrically.
In any event, of all point-caustic transits, only a fraction $\beta/\rho$
have impact parameters smaller than $\beta$.

Third, binary-lens events are probably more common than 
point-mass caustic crossings.  The fraction of events with 
binary-caustic crossings
has not yet been established empirically, but 5\% appears to
be a plausible estimate.  The fraction of point-mass caustic crossings
is $\langle \rho\rangle$, but by the argument of the previous paragraph,
only about half of these are really useful.  The mean radius of a giant is 
 $R\sim 22\,R_\odot$ (Gould 1995b), 
about 2.2 times larger than the fiducial value
used in equation (4.1).  Thus, the fraction of events with useful point-mass
caustic crossings is $\sim 0.5\langle \rho\rangle\sim 2\%$.

On the other hand, as shown by Figure 4, over the range probed by the
point lens ($r<\beta$), the point-lens errors are less than half as large
as those of the binary.  This advantage diminishes as $\rho^{-1/4}$ for
larger stars, but is still substantial for most giants.  The problem
of recognizing events with $\beta\ll\rho$ sufficiently early to
permit spectroscopic monitoring is formidable.  However, if they can
be monitored beginning at their peak, they are the best events to follow.

\begin{acknowledgements}
 
This work was supported in part by grant AST 94-20746 from the NSF.  

\end{acknowledgements}

%-------------------------------------------------------------------------
\newpage
\begin{figure}
\postscript{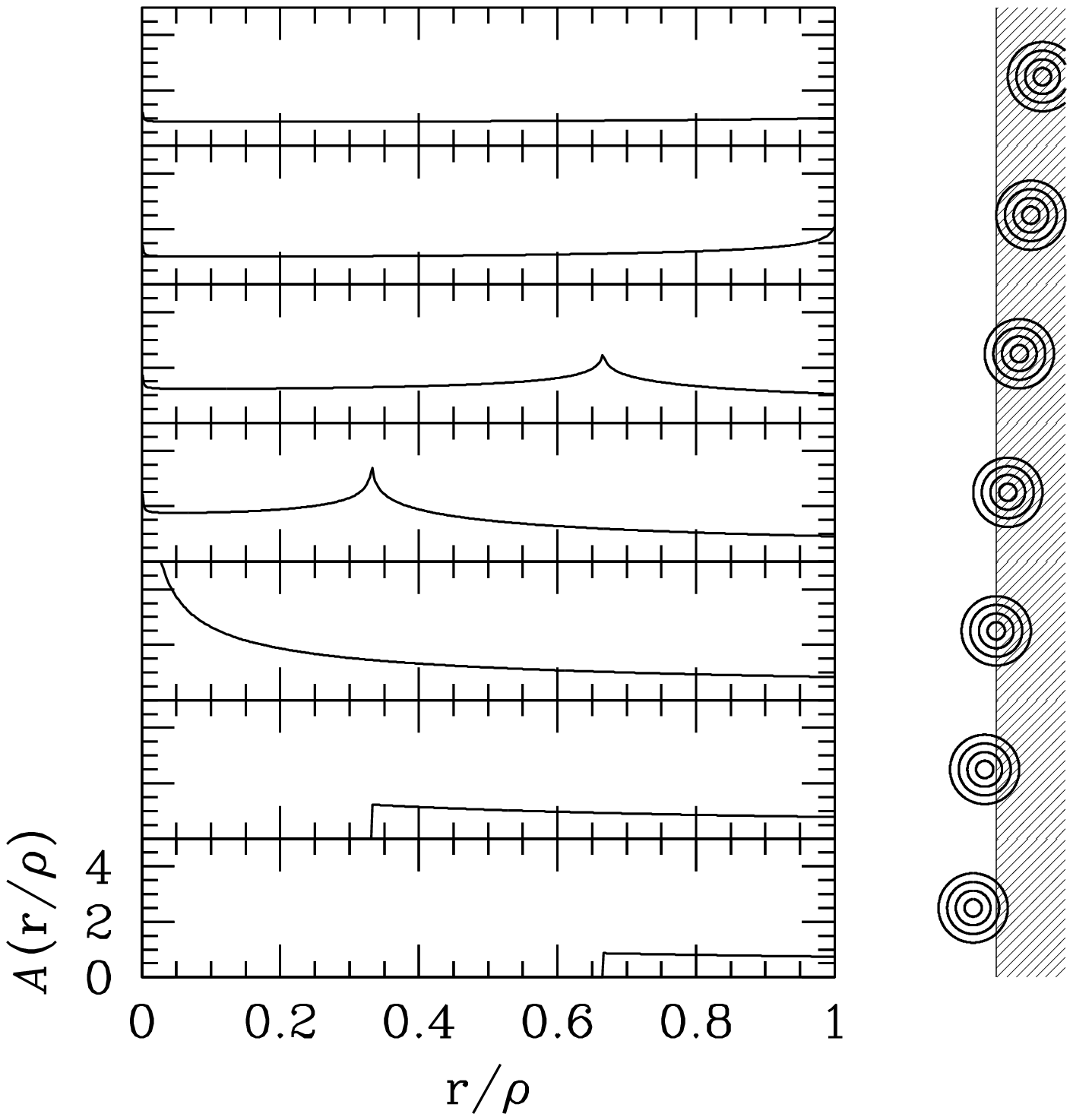}{1.1}
\caption{
The angle averaged magnification, ${\cal{A}}$ as a function of the
radius in units of the source radius $r/\rho$ for the fold caustic
case, for seven different times, 
$x=-t/t_e=5/3\rho$ (top panel),
 $\rho$, $2/3\rho$, $1/3\rho$, $0$, $-1/3\rho$, and
 $-2/3\rho$ (bottom panel).  
The associated geometry is shown to the right of
each panel.  We have set the magnification not associated with the caustic
to zero, and have taken $\rho=1.0$.  For other source radii,
 ${\cal{A}}$ scales as $\rho^{-1/2}$.
}

\end{figure}
\newpage

\begin{figure}
\postscript{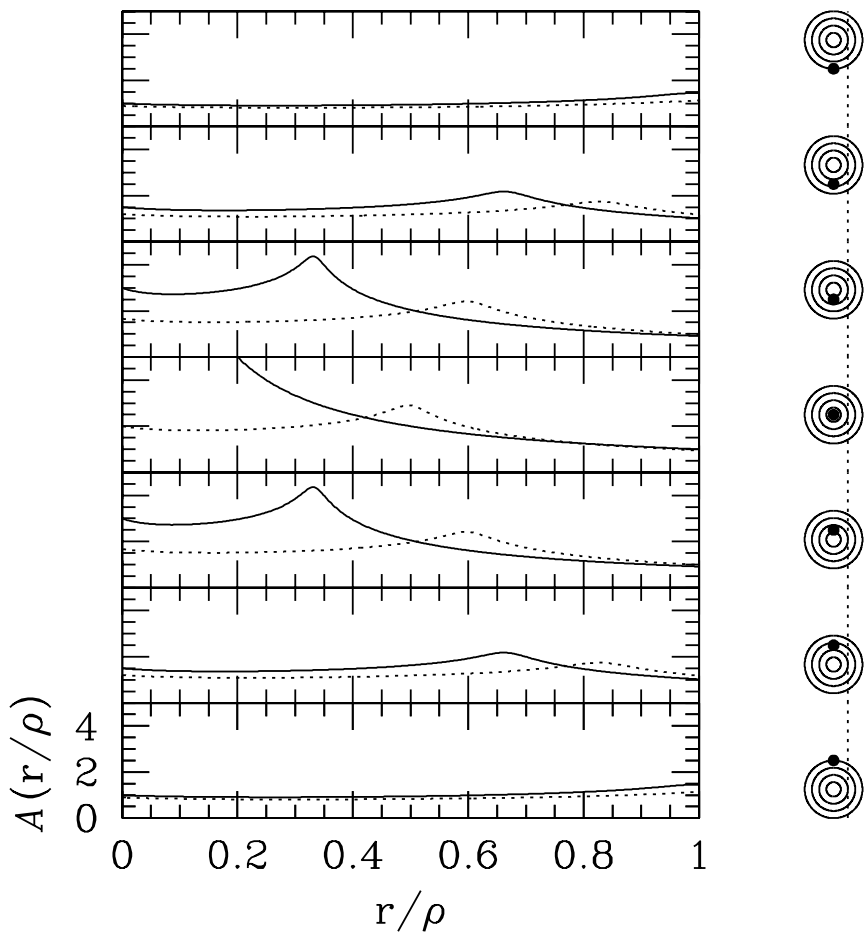}{1.1}
\caption{
The angle averaged magnification, ${\cal{A}}$ as a function of the
radius in units of the source radius $r/\rho$ for the point caustic
case, for seven different times, 
$t/t_e=-\rho$ (top panel),
 $-2/3\rho$, $-1/3\rho$, $0$, $1/3\rho$, $2/3\rho$, and
 $\rho$ (bottom panel), 
and two different trajectories, $\beta=0$ (solid
 lines) and $\beta=0.5\rho$ (dashed lines).  The associated geometry is shown to the right of
each panel.  We have taken $\rho=1.0$.  For other source radii,
 ${\cal{A}}$ scales as $\rho^{-1}$.
}
\end{figure}
\newpage
\begin{figure}
\postscript{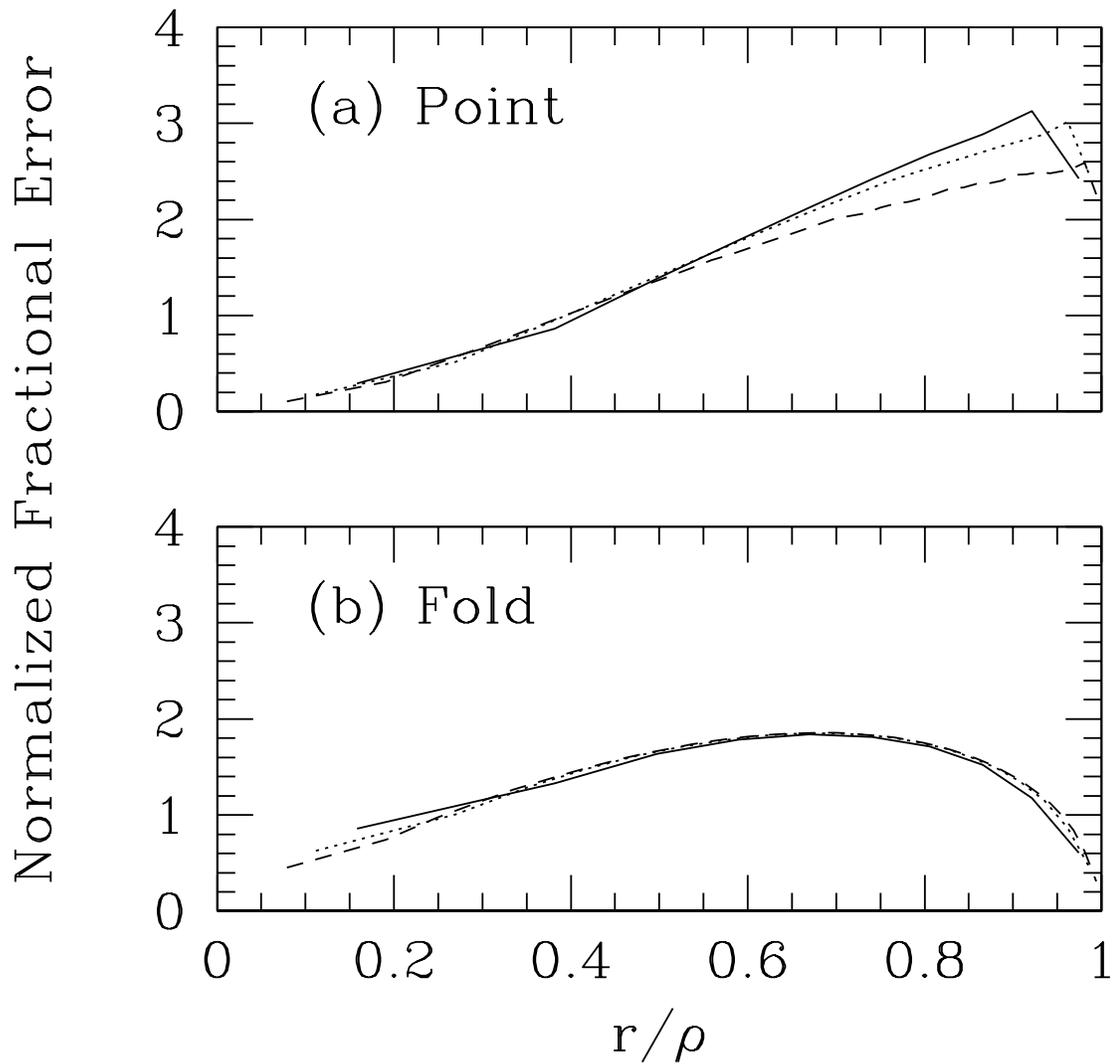}{1.1}
\caption{
(a) The normalized fractional error in the recovered intensity profile
 ($\delta I/I$) as a function of the radius in units of the source
 radius $r/\rho$ for the point caustic.
Each line corresponds to a different number of radial bins $N_r$ in
the recovered intensity profile, $N_r=10$ (solid), $20$ (dotted), 
and $40$ (dashed).  (b) Same as (a) except for the fold caustic. 
}
\end{figure}
\newpage
\begin{figure}
\postscript{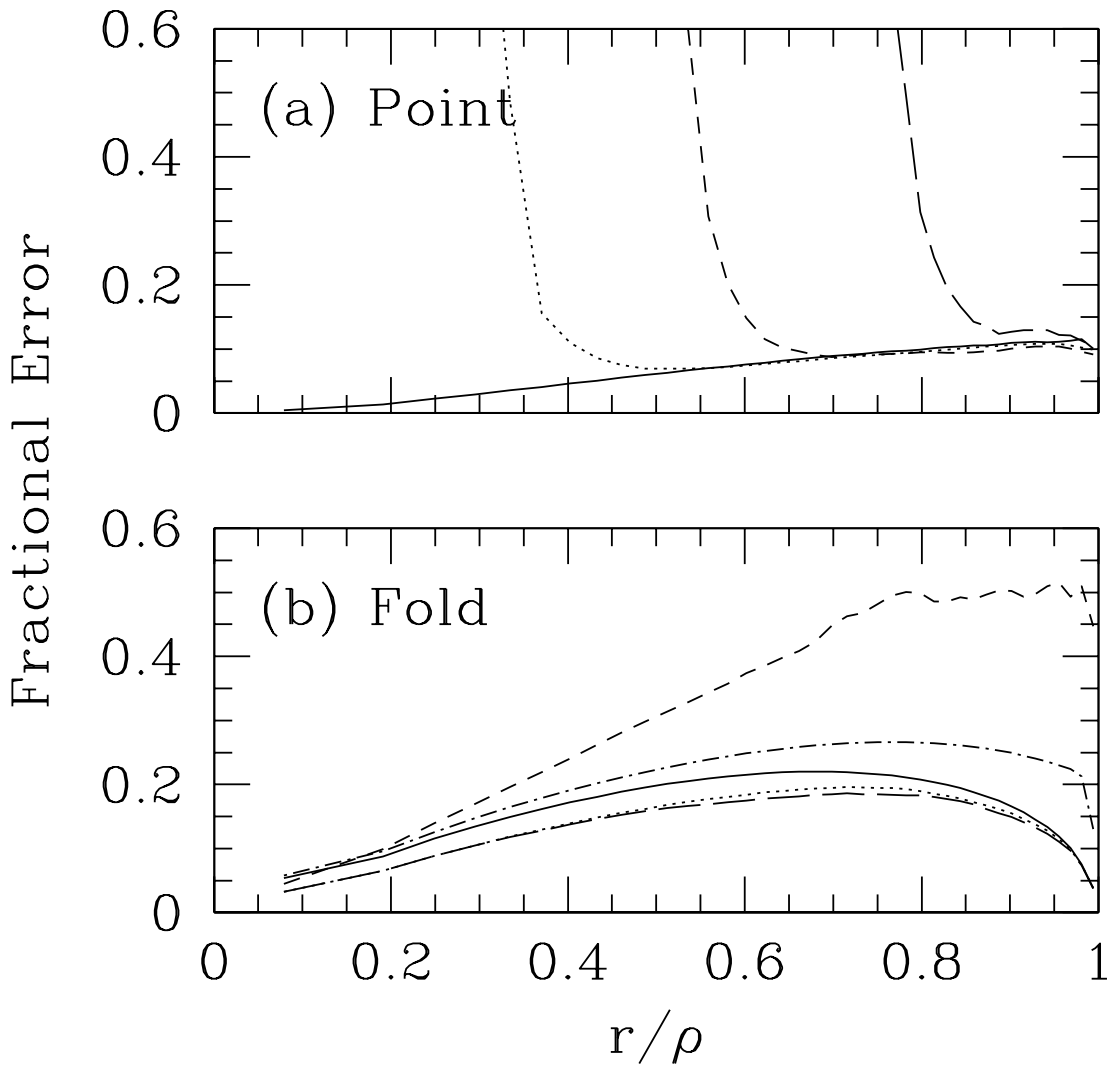}{1.1}
\caption{
The fractional error in the recovered intensity profile ($\delta I/I$) as a function of the
radius in units of the source radius $r/\rho$ for the fiducial
parameters: $N_r=10$ (number of radial bins in the recovered intensity profile), 
 $\gamma=0.4\, {\rm{s^{-1}}}$ (unlensed flux of star),
 $t_{\cross}=7\,{\rm hours}$ (crossing time of star) and $\rho=0.02$. 
(a) The errors for the point caustic,
for $t/t_e$ from $0$ to $\rho$ and four different values of the impact
parameter, $\beta/\rho=0$ (solid), $0.25$ (dotted), $0.5$ (short
dashed), and $0.75$ (long dashed).
(b) The errors for the fold caustic,
for four different trajectories: 
$x=-t/t_e$ from $0$ to $-\rho$ (solid),
 $\rho$ to $0$ (short dashed), $\rho$ to $-\rho$ (dotted), and $2\rho$ to
 $-\rho$ (long dashed). 
For these curves, we have set the
magnification not associated with the
 caustic to zero ($A_0=0$).  We also show the fractional error for the
 case $A_0=1$ and 
 $x=-t/t_e$ from $0$ to $-\rho$ (dashed-dot).
}
\end{figure}
\newpage
\begin{figure}
\postscript{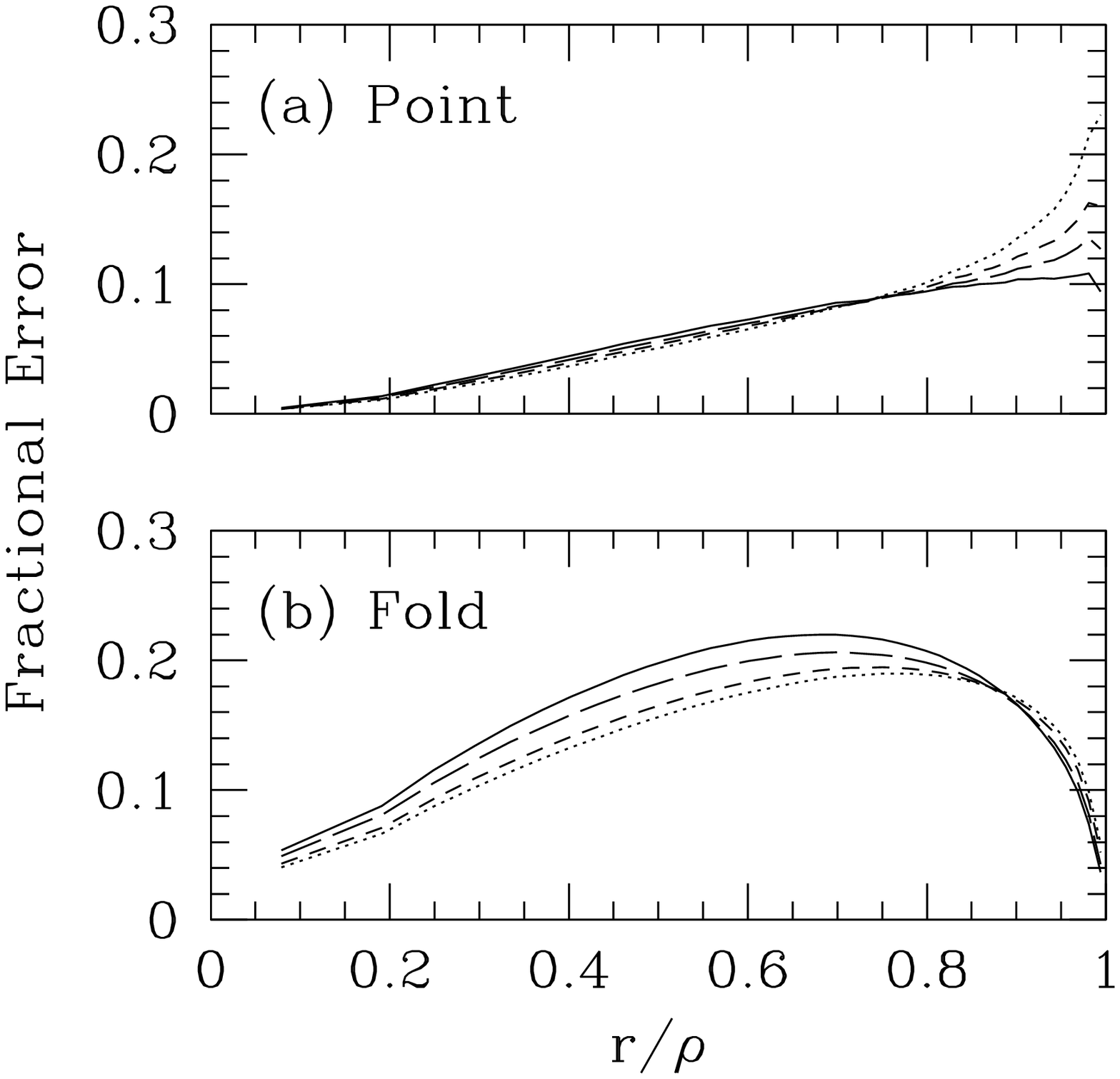}{1.1}
\caption{
(a) The fractional error in the recovered intensity profile ($\delta
I/I$) as a function of the
radius in units of the source radius, $r/\rho$, for the point caustic,
for a uniform source (solid), and limb-darkened source
as appropriate for
the broad bands $V$ (dotted), $I$ (short dashed), and $K$ (long dashed).
(b) Same as (a) except for the fold caustic.  For both (a) and (b) we
have assumed the same fiducial parameters as the solid curves in
Figure 4. 
}
\end{figure}

\newpage
\begin{references}

\reference{Albrow et al.\ 1997, in Variable Stars and the
  Astrophysical Returns of Microlensing Surveys, eds. R. Ferlet,
  J.-P. Maillard, \& B. Raban (Gif-sur-Yvette: Editions Fronti\'eres),
  135}

\reference{Albrow, M.\ et al.\ 1998a, ApJ, submitted}

\reference{Albrow, M.\ et al.\ 1998b, in preparation}

\reference{Alcock, C.\ et al.\ 1997a, ApJ, 479, 119} %bulge

\reference{Alcock, C.\  et al.\ 1997b, ApJ, 491, 436} %source res.

\reference{Ansari, R.\ et al.\ 1996, A\&A, 314, 94}

\reference{Gould, A.\ 1994, ApJ, 421, L71}

\reference{Gould, A.\ 1995a, ApJ, 444, 556}

\reference{Gould, A.\ 1995b, ApJ, 447, 491}

\reference{Gould, A.\ 1996, PASP, 108, 465}

\reference{Gould, A.\ 1997, ApJ, 483, 98}

\reference{Gould, A., \& Gaudi, B.S.\ 1997, ApJ, 486, 687}

\reference{Gould, A., \& Loeb, A.\ 1992, ApJ, 396, 104}

\reference{Gould, A., \& Welch, D.\ 1996, ApJ, 464, 212}

\reference{Griest, K. et al.\ 1991, ApJ, 372, 79}

\reference{Heyrovsky, D., Sasselov, D., \& Loeb, A.\ 1998, in preparation}

\reference{Lennon, D.J. et al.\ 1997, The Messenger, in press}

\reference{Loeb, A., \& Sasselov, D.\ 1995, ApJ, 449, 33L}

\reference{Manduca, A.\ 1979, A\&AS, 36, 411}

\reference{Manduca, A., Bell, R.A., \& Gustafsson, B.\ 1977, A\&A, 61, 809}

\reference{Mao, S., \& Paczy\'nski, B.\ 1991, ApJ, 374, 37}

\reference{Maoz, D., \& Gould, A.\ 1994, ApJ, 425, L67}

\reference{Nemiroff, R. J., \& Wickramasinghe, W.A.D.T.\ 1994, ApJ, 424, L21}

\reference{Paczy\'nski, B.\ 1986, ApJ, 304, 1}

\reference{Paczy\'nski, B.\ 1991, ApJ, 371, 63}

\reference{Pratt, M.R. et al.\ 1996, in IAU Symp. 173, Astrophysical Applications
of Gravitational Microlensing, e. C.S. Kochanek \& J.N. Hewitt
(Dordrecht: Kluwer), 221}

\reference{Sasselov, D.\ 1996, in 12th IAP Astrophysics Colloquim, ed. R. Ferlet}

\reference{Schneider, P., Ehlers, J., \& Falco, E.E.\ 1992, Gravitational Lenses
(Berlin: Springer)}

\reference{Schneider, P., \& Weiss, A.\ 1987, A\&A, 171, 49}

\reference{Udalski, A., et al.\ 1997, Acta Astron., 47, 169}

\reference{Valls-Gabaud, D.\ 1994, in Large scale structures of the universe,
ed. J. Mucket et al.\ (World Scientific), 326}

\reference{Valls-Gabaud, D.\ 1997, MNRAS, submitted}

\reference{Witt, H.\ 1995, ApJ, 449, 42}

\reference{Witt, H., \& Mao, S.\ 1994, ApJ, 430, 505}

\reference{Witt, H., \& Mao, S.\ 1995, ApJ, 447, L105}




\end{references}
\end{document}